\documentclass[a4paper,aps,pre,floatfix,twocolumn,nofootinbib,showpacs,superscriptaddress]{revtex4-1}

\usepackage[pdftex]{graphicx}
\usepackage{latexsym,amsmath}
\usepackage[pdftex]{color}
\usepackage{hyperref}
\usepackage{float}
\usepackage[all]{xy}
\usepackage{verbatim}

\newcommand{\av}[1]{\langle #1 \rangle}

\begin{document}

\title{Aging and percolation dynamics in a Non-Poissonian temporal
  network model}

\author{Antoine Moinet}

\affiliation{Departament de F\'{\i}sica, Universitat Polit\`ecnica de
  Catalunya, Campus Nord B4, 08034 Barcelona, Spain}

\affiliation{Aix Marseille Universit\'{e}, Universit\'{e} de Toulon,
  CNRS, CPT, UMR 7332, 13288 Marseille, France}

\author{Michele Starnini}

\affiliation{Departament de F\'{\i}sica Fonamental, Universitat de
  Barcelona, Mart\'{\i} i Franqu\`es 1, 08028 Barcelona, Spain}

\author{Romualdo Pastor-Satorras}

\affiliation{Departament de F\'{\i}sica, Universitat Polit\`ecnica de
  Catalunya, Campus Nord B4, 08034 Barcelona, Spain}

\begin{abstract}
  We present an exhaustive mathematical analysis of the recently proposed Non-Poissonian Activity Driven (NoPAD) model 
  [Moinet  \textit{et al.} \textit{Phys. Rev. Lett.}, \textbf{114} (2015)], 
  a temporal network model incorporating the empirically observed bursty nature of social interactions. 
  We focus on the aging effects emerging from the Non-Poissonian dynamics of link activation,
  and on their effects on the topological properties of time-integrated networks, such as the degree distribution.  
  Analytic expressions for the degree distribution of integrated networks as a function of time are derived, 
  exploring both limits of vanishing and strong aging.
  We also address the percolation process occurring on these temporal networks, 
  by computing the threshold for the emergence of a giant connected component, highlighting the aging dependence. 
  Our analytic predictions are checked by means of extensive numerical simulations of the NoPAD model.
\end{abstract}


\maketitle

\section{Introduction}

The network science approach to complexity has been traditionally based
in a static view of complex systems \cite{Newman2010, Dorogobook2010}.
The increasing availability of time-resolved data on different kind of
interactions has unveiled an additional level of complexity in networked
systems, consisting in topological patterns of connections that evolve
with time \cite{Holme:2011fk,citeulike:13696788}. This transformation
has been particularly relevant in the field of social sciences
\cite{Jackson2010,lazer2009life}, since social interactions are
intrinsically dynamic, being constantly created and terminated at
different time scales.  Moreover, digital traces of human dynamics are
nowadays ubiquitous, from mobile phone communications \cite{Onnela:2007}
to face-to-face social interactions \cite{10.1371/journal.pone.0011596},
providing inestimable longitudinal data, including timing of social
interactions, at a scale unprecedented in other kinds of systems.  The
emerging field of temporal networks
\cite{Holme:2011fk,citeulike:13696788}, developed to yield the
theoretical grounding needed to represent and analyze the properties of
such time-varying complex systems, is thus influenced by a bias towards
social dynamics.

From the study of large scale data, a wealth of rich patterns and
characteristic properties have been uncovered in human dynamics.  One of
the most striking features observed is probably the bursty nature of
social interactions \cite{Barabasi:2005uq}, revealed by the observation
of inter-event times $\tau$ between two consecutive interactions of the
same individual following heavy tailed distributions that can
approximated as power laws of the form
$\psi(\tau) \sim \tau^{-(1+\alpha)}$, with $0 < \alpha < 2$ in general,
at strong odds with previously assumed Poissonian behavior. Such bursty
behavior has been observed in many instances of social interactions
\cite{Oliveira:2005fk,10.1371/journal.pone.0011596,holme2003} as well as
in phenomena belonging to other fields, such as earthquakes
\cite{citeulike:12015394}, neuronal activity
\cite{DBLP:journals/biosystems/KemuriyamaOSMTKN10}, mRNA synthesis in
cells \cite{citeulike:840929}, etc. (see Ref.~\cite{citeulike:13696788}
for an extensive reference list). Moreover, it has been recognized that
the bursty nature of temporal networks can have a deep impact on
dynamical processes running on top of such time-varying systems
\cite{burstylambiotte2013, Takaguchi2012, PhysRevE.83.025102,
  PhysRevE.85.056115, min_spreading_2011,perra_random_2012,
  DBLP:journals/corr/PerottiJHS14}. These observations claim for a
better theoretical understanding of the dynamics of such temporal
networks, and in particular for the design of simplified, descriptive or
generative temporal network models.

Several models of temporal networks have been put forward in the
literature
\cite{holme2013,starnini_modeling_2013,Mantzaris2012,PhysRevE.90.042805,Colman2015},
focused on different possible mechanisms to explain the empirically
observed properties. Among those, it is noteworthy the recently proposed
Non-Poissoinan activity driven (NoPAD) model
\cite{PhysRevLett.114.108701}. The NoPAD model aims to be a generative
model, akin to the configuration model for static networks
\cite{Bender1978}. It is defined in terms of agents that follow
independent renewal processes \cite{renewal}, starting social
connections separated by intervals of time $\tau$ distributed according
to some prescribed waiting time distribution $\psi(t)$, establishing
connections to randomly chosen other agents. The model thus captures the
most basic feature of social temporal networks, namely a non Poissonian
inter-event time distribution, which can be adjusted by imposing long
tailed waiting time distributions $\psi(t)$, in a simple, mathematically
tractable framework, the one of renewal theory \cite{renewal}.

In this paper we present a detailed mathematical analysis of the NoPAD
model, focusing in the properties of the static networks that can be
constructed integrating the contacts in the temporal dynamics.  Indeed,
within the mathematical framework of temporal network, a static
representation can be recovered by integrating a time-varying graph in a
time interval $[t_a , t_a + t]$, spanning a width $t$, and starting
after a time $t_a$ from the inception of the network. The study of this
integrated network is relevant, since traditional static social networks
\cite{Jackson2010} are constructed in this way, and it is important to
know how the features of the temporal dynamics affects the topological
properties of its integrated counterpart.  The inclusion of
non-Poissoinian dynamics in the process of links addition, given by the
waiting time $\psi(t)$ with a non-exponential form, has a deep impact on
the topology of the resulting time-aggregated network.  A relevant
signature of this effect is the aging behavior \cite{Henkel2,
  klafter_first_2011} of its topological properties, which depends not
only on the width of the integration time $t$
\cite{ribeiro_quantifying_2013}, but also on the \textit{aging time}
$t_a$ at which the integration starts. The NoPAD model can thus be
viewed as a null model, able to single out aging exclusively due to the
burstiness of links activation, different from aging of different nature
that might be present in real social networks, such as physical aging
\cite{medo_temporal_2011, zhu_effect_2003}.

The paper is structured as follows: In Section \ref{sec:non-poiss-activ}
we define the NoPAD model as a natural extension of the previously
proposed activity driven model
\cite{2012arXiv1203.5351P}. Sec.~\ref{sec:topol-prop-integr} sets up the
mapping of the NoPAD model to a hidden variables network
\cite{PhysRevE.68.036112}, which will allow for the calculation of the
topological properties of the integrated network.
Sec.~\ref{sec:degree-distribution} is devoted to the general calculation
of the integrated degree distribution. The case of waiting time
distributions with a power-law form in the absence of aging, i.e. with
$t_a=0$, is described in Sec.~\ref{sec:non-aged-networks}; the more
interesting case of aging effects is analyzed in
Sec.~\ref{sec:aged}. The percolation properties of the
integrated network are discussed in Sec.~\ref{sec:perco}, where we study
the time $T_p$ at which a giant component in the integrated network
emerges, spanning a finite fraction of the network.  Finally,
Section~\ref{sec:conclusions} concludes the paper, discussing the
results presented and drawing future perspectives.

\section{The Non-Poissonian Activity Driven Model}
\label{sec:non-poiss-activ}

The activity-driven (AD) model \cite{2012arXiv1203.5351P} is built upon
the empirical observation that individuals are characterized by
different levels of \textit{social activity}, i.e. they have different
levels of propensity to become engaged in social interactions. 
Social activity, which can be defined as the probability per unit time $a$ 
that an individual becomes active and starts a social interaction, 
has been empirically measured in a variety of social temporal networks, 
and shown to exhibit a heterogeneous, heavy tailed distribution \cite{2012arXiv1203.5351P}.  
To take into account this heterogeneity, in the AD model
\cite{2012arXiv1203.5351P,starnini_topological_2013} each node $i$,
representing an agent, is endowed with a \textit{constant} activity $a_i$, 
representing the probability that at each time step he/she will establish a link, 
of infinitesimally short duration, with another agent, chosen uniformly at random.  
It is possible to show \cite{2012arXiv1203.5351P,starnini_topological_2013} that 
the degree distribution $P_t(k)$ of the resulting network integrated up to time $t$
is functionally related to the probability distribution $F(a)$ from which the activities $a_i$ are drawn.  
Therefore, if fed with the empirically observed $F(a)$, the time-integrated AD networks show 
some of the topological properties of real social networks, 
and in particular its characteristic heavy tailed degree distribution \cite{starnini_topological_2013}.  
The AD model has proved to be very flexible, allowing to incorporate many typical features of human dynamics, 
such as memory effects \cite{Karsai:2014ab}, and it is analytically suitable to study dynamical processes on 
time-varying networks, such as epidemic spreading \cite{PhysRevLett.112.118702},
random walks \cite{perra_random_2012,glassangelica}, or percolation \cite{citeulike:12856686}.

However, it is easy to see that for sufficiently large $N$, in the continuous time limit, 
a constant activity $a_i$ leads to an inter-event time distribution for node $i$ 
with the form $\psi_i(\tau) = a_i e^{-a_i \tau}$ \cite{glassangelica}, following an exponential form.
Thus the AD model fails to reproduce the main feature observed in real temporal networks, 
namely a long tailed inter-event time distribution between social contacts.
One way to overcome this drawback is to allow for each node $i$ a time-dependent activity $a_i(\tau)$, 
where $\tau$ is the time elapsed since the last activation of node $i$. 
With this assumption, one can define a generalized model in which each individual $i$ becomes active 
by following a renewal process \cite{renewal}, 
defined by a \textit{waiting time} distribution between successive activation events 
given by $\psi_i(\tau) = a_i(\tau) \exp \{ - \int_0^{\tau} a_i(t) dt \}$ \cite{renewal}. 
For the standard AD model, with $a_i$ constant, we have $\psi_i^\mathrm{AD}(\tau) = a_i e^{- a_i \tau}$, 
that is, agents follow a simple Poisson process \cite{kingman1992poisson}. 
Any function $a_i(\tau)$ leads thus to the consideration of a generalized 
\textit{non-Poissonan activity driven} (NoPAD) model \cite{PhysRevLett.114.108701}. 
Shifting away from the instantaneous activity $a_i(\tau)$, the NoPAD model can be defined in terms of a set
of agents that become active by following a renewal process with waiting time distribution $\psi_i(\tau)$, 
giving the probability of observing a time $\tau$ between two activation events of agent $i$. 
For the sake of simplicity, we assume here the same functional form of $\psi_i$ for all agents, 
$\psi_i(\tau) \equiv \psi_{c_i}(\tau)$, where the parameter $c$ gauges the heterogeneity of the activation rate of the agents, 
and it is randomly drawn form a distribution $\eta(c)$.

Here we are interested in reconstructing the integrated network obtained by aggregating interactions 
occurring within the time interval $[t_a, t_a+t]$. 
in order to build such networks in the non-aged case, $t_a = 0$, we proceed as follows:
\begin{itemize}
\item We start with a set of $N$ disconnected nodes.
\item For each agent $i$, we extract $r_i+1$ waiting times $\tau_k$,
  with $k = 1,2, \ldots, r_i+1$ from the probability distribution
  $\psi_{c_i}(\tau)$, and define the activation times
  $T_{j} = \sum_{k=1}^{j} \tau_k$, such that $T_{r_i+1} >t$ and
  $T_{r_i}<t$.  In this way, individual $i$ has been active $r_i$ times
  within the interval $[0,t]$.
\item Each time an agent $i$ is active, an individual $j \neq i$ is chosen uniformly at random 
  and an edge is created between $i$ and $j$.  
  In the case of a pre-existing link, no additional edge is created (a weight increment may possibly be considered).
\end{itemize}
In order to construct the aged network, i.e. aggregated over the time
interval $[t_a,t_a+t]$, we apply the generating process between $0$ and
$t_a+t$ and discard all the events occurring before $t_a$.

\section{Mapping to hidden the variable formalism}
\label{sec:topol-prop-integr}

The topological properties of the integrated networks generated by the NoPAD model 
can be worked out by applying a mapping to the class of network models with hidden variables, 
proposed in Ref. \cite{PhysRevE.68.036112} (see also \cite{PhysRevLett.89.258702,Soderberg:2002fk}).  
Hidden variables network models are defined as follows: 
starting from a set of $N$ initially disconnected nodes, 
each node $i$ has assigned a variable $\vec{h}_i$, drawn at random from a probability distribution $\rho(\vec{h})$.  
Each pair of nodes $i$ and $j$, with hidden variables $\vec{h}_i$ and $\vec{h}_j$, 
are connected with an undirected edge with probability $\Pi(\vec{h}_i, \vec{h}_j)$ (the connection probability).
The model is fully defined by the functions $\rho(\vec{h})$ and $\Pi(\vec{h}, \vec{h}')$, 
and all the topological properties of the resulting network can be derived through 
the propagator $g(k|\vec{h})$ \cite{PhysRevE.68.036112},
defined as the conditional probability that a vertex with hidden variable $\vec{h}$ ends up connected to 
exactly $k$ other vertices  (has degree $k$).  
From this propagator, expressions for the topological properties of the model 
can be readily obtained \cite{PhysRevE.68.036112}.

We can apply the hidden variables formalism to the NoPAD model defined in Sec.~\ref{sec:non-poiss-activ} 
by identifying the mapping to the corresponding hidden variables and connection probability. 
Let us assume that all agents are disconnected and synchronized at time $t=0$, 
and let us consider an integration time window $[t_a, t_a + t]$. 
From the definition of the model, the parameter that determines the connectivity of a node $i$ is 
the number of times $r_i$ that it has become active in the considered time window (its activation number). 
This number depends on its turn of the parameter $c_i$ characterizing the waiting time
distribution of node $i$. Therefore, we choose as hidden variables
\begin{equation}
  \vec{h} \to (r, c).
\end{equation}
It is worth noticing that these quantities are not independent, and it is convenient to describe the variable $r$ 
with its conditional distribution $\chi_{t_a, t}(r \vert c)$, 
which can be computed in terms of $\psi_c(\tau)$ \cite{klafter_first_2011}.  
The hidden variable probability distribution thus reads
\begin{equation}
  \rho(\vec{h}) \to \rho_{t_a, t}(r,c) \equiv \eta(c) \chi_{t_a, t}(r
  \vert c). 
\end{equation}
Finally, it is easy to see that the connection probability only depends on the activation numbers $r_i$, 
\begin{equation}
  \Pi(\vec{h},\vec{h'}) \to \Pi(r, r'),
\end{equation}
depending only implicitly on the integration window through the
distribution of the activation numbers $\chi_{t_a, t}(r \vert c)$. 
Simple probabilistic arguments show that the probability that two nodes with activation numbers $r$ and $r'$ 
become eventually connected in the integrated network is
$\Pi(r, r') = 1 - [1-N^{-1}]^{r + r'}$~\cite{starnini_topological_2013}. 
Thus, in the limit $N\gg r \,, r'$, we have
\begin{equation}
   \Pi(r, r')  \simeq \frac{r + r'}{N}.
   \label{eq:5}
\end{equation}

Given the form of the connection probability, 
the corresponding propagator will be a function of the activation number alone, $g(k | r)$. 
To find its functional form, one notices that a node with activation number $r$ will have a degree $k$ equal to 
the sum of an in-degree and and out-degree, $k = k_\mathrm{out} +k_\mathrm{in}$, 
which accounts for the edges created by the activation of the node, 
and by the activation of all other nodes, respectively. 
In the limit $N\gg r \,, r'$, $k_\mathrm{out} = r$ 
(each activation event leads to an edge connecting to a different node) and thus 
the propagator of the out-degree is a delta function centered at $r$, $g_\mathrm{out} (k|r) = \delta(k-r)$. 
For the in-degree we can write $k_\mathrm{in} = \sum_{r'}k_\mathrm{in}(r')$, 
where $k_\mathrm{in}(r')$ is the number of connections received from other nodes with hidden variable $r'$. 
By following \cite{PhysRevE.68.036112} we obtain that the generating function of the in-degree propagator,
$\hat{g}_\mathrm{in}(z|r) = \sum_k g_\mathrm{in}(k|r) z^k$, fulfills the equation
\begin{equation}
  \ln \hat{g}_\mathrm{in}(z|r) = N \sum_{r',c'} \rho_{t_a,t}(r',c')  \ln
  \left[  1-(1-z) \pi(r')\right] 
\end{equation} 
where $\pi(r') = 1-(1-1/N)^{r'}$ is the probability that a given node is reached at least once 
by a node with activation number $r'$.
Therefore, for a sparse network, the in-degree propagator reads \cite{PhysRevE.68.036112}
\begin{equation}
  g_\mathrm{in}(k \vert r) = e^{-\av{r}_{t_a,t}} \,\dfrac{(
    \av{r}_{t_a,t})^{k}}{k!}. 
\end{equation}
where we have defined the moments of the activation number distribution
$\av{r^n}_{t_a,t}=\sum_{r,c}\,r^n\,\eta(c) \chi_{t_a, t}(r \vert c)$ 
in the time window $[t_a, t_a+t]$, which we shall from now on write as
$\av{r^n}$ (or $\av{r^n}_0$ when we explicit consider a non-aged process) 
for the sake of legibility.  Finally, we obtain the total propagator as 
the convolution of the in-degree and the out-degree propagators, having the form
\begin{equation}
g(k\vert r) = \left\{
    \begin{array}{lr}
      e^{-\av{r}}\,\frac{\av{r}^{k-r}}{(k-r)!} & \mathrm{for}  \; k\geq r\\
      0 & \mathrm{otherwise}
    \end{array} \right. .
    \label{eq:gk}
\end{equation}
In the limit $\av{r} \gg 1$, the previous exact expression can be
approximated by the simple shifted Poissonian form
\begin{equation}
  g(k\vert r)=e^{-(r+\av{r})}\,\dfrac{(r+\av{r})^k}{k!},
  \label{eq:31}
\end{equation}
which we will use in the rest of the manuscript to allow for mathematical tractability.

\section{General form of the degree distribution}
\label{sec:degree-distribution}

The most relevant topological property of any static network is its degree distribution $P(k)$, 
defined as the probability that a randomly chosen node has degree $k$ \cite{Newman2010}. 
The degree distribution generated by the NoPAD model in an integration window $[t_a, t_a+t]$ 
can be expressed in terms of the propagator $g(k\vert r)$ as \cite{PhysRevE.68.036112}
\begin{equation}
  \label{eq:pk}
  P_{t_a, t}(k) = \sum_{c, r} \rho_{t_a, t}(r,c)  g(k\vert r) . 
\end{equation}
The general asymptotic form of the degree distribution can be obtained by
performing a steepest descent approximation.  For $\av{r} \, \gg 1$,
using the Poissonian propagator Eq.~(\ref{eq:31}), and considering $r$
as a continuous variable, we can write Eq.~(\ref{eq:pk}) as
\begin{equation}
  \label{eq:pk_psi}
  P_{t_a, t}(k) \simeq \sum_c \eta(c) \int dr \,
  e^{\phi(r)}\chi_{t_a,t}(r\vert c) 
\end{equation}
where 
\begin{equation}
  \phi(r)=-\av{r}-r+ k\, \ln(r+\av{r})-\ln(k!) .
\end{equation}
This function has a maximum at $r_m = k-\av{r}$ and its second
derivative at this point is $\phi^{''}(r_m)=-\frac{1}{k}$.  By expanding
$\phi$ up to second order, one can obtain
\begin{equation}
  e^{\phi(r)} \simeq  \dfrac{e^{-(r-r_m)^2/2k}}{\sqrt{2 \pi k}}  \simeq
  \delta(r-r_m), 
\end{equation}
where we have used Stirling's approximation, and replaced the ensuing
Gaussian function by a Dirac delta function.
Therefore, the degree distribution reads
\begin{equation}
  P_{t_a, t}(k) \simeq \sum_c \eta(c)\, \chi_{t_a,t}(r_m \vert c),
\label{eq:7}
\end{equation}
where we recall $r_m = k-\av{r}$.

If we consider the simple case of a Poissonian inter-event time distribution, $\psi(\tau) = c e^{-c \tau}$, 
as in the original AD model, the activation number distribution is simply given by the Poisson distribution \cite{renewal},
\begin{equation}
\chi_{t_a,t}(r\vert c) = e^{-ct}\dfrac{(ct)^r}{r!},
\end{equation}
with an average activation number $\av{r}=\av{c}t$,  which is independent of the aging time $t_a$ 
due to the memoryless nature of the Poisson process \cite{renewal}.  
In a continuous $c$ approximation, defining $\chi_{t_a,t}(r_m\vert c)=e^{\varphi(c)}$ with
\begin{equation}
  \varphi(c) = -ct+r_m\mathrm{ln}(ct)-\mathrm{ln}(r_m!)
\end{equation}
and applying once again a steepest descent approximation around the maximum at $c_m=\frac{r_m}{t}$, 
with the condition $\vert \varphi^{''}(c_m)\vert =\frac{t^2}{r_m} \gg 1$, one finally obtains
\begin{equation}
  P_t(k) 
  \simeq \dfrac{1}{t}\eta\left(\dfrac{k}{t}-\av{c}\right),
\label{eq:27}
\end{equation}
recovering the asymptotic form of the integrated degree distribution obtained in \cite{starnini_topological_2013}, 
the limits of its validity being $\av{c}t \gg 1$, and $t^2 \gg (k-\av{c}t) \gg 1$.

In the case of heavy tailed waiting time distributions, we expect to
observe strong departures form the simple result in Eq.~(\ref{eq:27}). 
Let us focus in particular on the simple power law form
\begin{equation}
  \psi_c(t) = \alpha \, c'
  \left(c'\,t+1\right)^{-(\alpha+1)},
  \label{eq:14}
\end{equation}
where $c' = c \, (\Gamma_{1-\alpha})^{\frac{1}{\alpha}}$, $c$ being the
parameter quantifying the (possible) heterogeneity of waiting times in
the population, and where we have defined $\Gamma_{z} \equiv \Gamma(z)$,
the gamma function.  We will explore in particular the case
$0 < \alpha < 1$, for which the first moment of the waiting time
distribution diverges, and aging effects are expected to be relevant
\cite{aging}.  We will work out approximations at large time; however,
since we used the sparse network hypothesis to obtain Eq.~(\ref{eq:31}),
we shall check afterwards that the condition $\frac{\av{r}}{N}\ll 1$
remains fulfilled.

As we can see from Eq.~(\ref{eq:7}), the degree distribution
$P_{t_a, t}(k)$ depends mainly on the activation time distribution
$\chi_{t_a, t}(r \vert c)$. Expressions for this function in the case of
a general waiting distribution $\psi_c(\tau)$ can be obtained in Laplace
space. Thus, defining the Laplace transforms
\begin{eqnarray}
  \psi_c(s) & = & \int_0^\infty d\tau \; \psi_c(\tau)  e^{-\tau s},
                  \nonumber \\ 
  \chi_{u, s}(r \vert c) & = & \int_0^\infty dt_a \; \int_0^\infty dt \;
  \chi_{t_a, t}(r \vert c) e^{-u t_a} e^{- s t}, \nonumber
\end{eqnarray}
we have~\cite{godreche,Barkai2003,aging}
\begin{equation}
  \chi_{u, s}(r \vert c) = \left\{
    \begin{array}{lr}
      (u s)^{-1} - h_c(u, s) s^{-1} &  \; r=0\\
      h_c(u, s) \psi_c(s)^{r-1} [1-\psi_c(s)]s^{-1} &   \;
                                                      r\geq 1\\
    \end{array} \right. ,
    \label{eq:chi_lapl}
\end{equation}
where
\begin{equation}
  h_c(u, s) = \frac{\psi_c(u) - \psi_c(s)}{s-u}
  \frac{1}{1-\psi_c(u)}
  \label{eq:2}
\end{equation}
is the double Laplace transform of the forward waiting time distribution
$h_c(t_a, t)$, defined as the distribution of the waiting time measured
forward from an arbitrary $t_a$ to the next activation of an individual.

\section{Non-aged networks}
\label{sec:non-aged-networks}

In the case of non-aged networks, it is easy to verify that
$h_c(t_a=0,t)=\psi_c(t)$, and Eq.~\eqref{eq:chi_lapl} reduces thus to
\begin{equation}
  \chi_s(r\vert c)= \psi_c(s)^{r}\dfrac{1-\psi_c(s)}{s},
  \label{eq:4}
\end{equation}
where the activation number distribution $\chi_t(r\vert c)$ depends now
only on the window length $t$.  By virtue of a Tauberian theorem
\cite{WeissRandomWalk}, for $\frac{s}{c} \ll 1$ we can expand
\begin{equation}
  \psi_c(s) \simeq 1-\left(\dfrac{s}{c}\right)^{\alpha},
  \label{eq:9}
\end{equation}
and using Eq.~\eqref{eq:4}, we deduce \cite{aging}
\begin{equation}
  \chi_t(r \vert c) \simeq \frac{1}{\alpha \, r}\ \,
  \dfrac{c\,t}{r^{1/\alpha}} \; \mathcal{L}
  \left(\dfrac{c\,t}{r^{1/\alpha}}\right) ,
\end{equation}
valid for $ct \gg 1$, and where $\mathcal{L}(z)$ is a one-sided L\'evy distribution 
with Laplace transform $\mathcal{L}(s) = e^{-s^{\alpha}}$ \cite{klafter_first_2011}. 
Using the expansion at large
$r$~\cite{godreche}
\begin{equation}
  \chi_t(r\vert c) \simeq
  \dfrac{(c\,t)^{-\alpha}}{\Gamma_{1-\alpha}}\exp\left(-(1-\alpha)
    \left[\left(\dfrac{\alpha}{ct}\right)^{\alpha}r
    \right]^{\dfrac{1}{1-\alpha}}\right) , \nonumber
\end{equation}
and inserting it into Eq.~\eqref{eq:7}, we arrive at
\begin{equation}
  \label{eq:pk_0}
  P_{t}(k) \simeq 
  \dfrac{(k-\av{r}_0)^{\frac{1}{\alpha}-1}}{\Gamma_{1-\alpha}t}\int
  \eta\left(\dfrac{u}{t}\,(k-\av{r}_0)^{\frac{1}{\alpha}}\right)
  \dfrac{e^{\xi(\alpha,u)}}{u^{\alpha}}    \, du ,
\end{equation}
where we have considered $c$ continuous, defined
$\xi(\alpha,u)=-(1-\alpha)\left(\alpha/u
\right)^{\frac{\alpha}{1-\alpha}}$, and $\av{r}_0$ is the average
activation number with no aging, $\av{r}_0 = \sum_c \eta(c) \sum_r r \chi_t(r\vert c)$.

Expression Eq.~(\ref{eq:pk_0}) depends now only on the waiting time heterogeneity distribution $\eta(c)$. 
While the parameter $c$ of an agent is not directly accessible from empirical data, 
in Ref.~\cite{PhysRevLett.114.108701} it was argued that it is directly related to the average activity $\bar{a}$, 
defined as the probability to become active in a time window of a given length $\Delta t$. 
Given the power law activity distribution observed in real temporal networks
\cite{2012arXiv1203.5351P}, here we assume a distribution
\begin{equation}
  \eta(c)=\dfrac{\beta}{c_0}\left(\dfrac{c}{c_0}\right)^{-(\beta+1)}, \;\;
  \beta > 0, \;\;c>c_0. 
\label{eq:eta_c}
\end{equation}
With this form of $\eta(c)$, the average activation number with no aging takes the form,
for large $t$ \cite{godreche},
\begin{equation}
  \av{r}_0 \simeq \frac{\beta \,\Gamma_{1+\alpha}^{-1}}{(\beta-\alpha)} (c_0 t)^\alpha,
\end{equation}
and the integral in Eq.~\eqref{eq:pk_0} has a lower bound at
$u_0 = c_0\,t\,(k-\av{r}_0)^{-\frac{1}{\alpha}}$. Taking the limit
$k-\av{r}_0 \;\gg (c_0 t)^{\alpha}$ we finally obtain the asymptotic
result
\begin{equation}
  P_t(k) \sim (c_0 t)^{\beta}(k-\av{r}_0)^{-\gamma}
\label{eq:16}
\end{equation}
with $\gamma = 1+\frac{\beta}{\alpha}$, an expression recovering the
analytic result obtained in Ref.~\cite{PhysRevLett.114.108701}, where it
was numerically confirmed by simulations of the NoPAD model.

The result Eq.~(\ref{eq:16}) is noteworthy in two respects. Firstly, it
relates two fundamental features found in real social networks, a broad
tailed inter-event time distribution, as represented by the wating time
distribution $\psi(\tau)$, and a scale free degree distribution
$P_t(k)$, whose exponent $\gamma$ is simply related to the parameters
$\alpha$, controlling $\psi(\tau)$, and $\beta$, related to the
heterogeneity of the individuals' social activity. Secondly, it shows
transparently that non-Markovian effects are related to an exponent
$\alpha<1$, associated with a diverging first moment of the waiting time
distribution. Indeed, in the limit $\alpha \to 1$, Eq.~(\ref{eq:16})
recovers the Poissonian results Eq. \eqref{eq:27}, which means that even
if the second moment of the waiting time distribution is infinite
($1<\alpha<2$), the structure of the integrated network will not be
significantly different (at dominant order in $t$) to that of a Poissonian AD network.

\section{Aging effects}
\label{sec:aged}

In the case $t_a >0$, and working in the large $t$ and $t_a$ limit, the
expression of the forward waiting time in Laplace space,
Eq.~(\ref{eq:2}), takes, by using Eq.~(\ref{eq:9}), the form
\begin{equation}
  h_c(u,s) = \dfrac{s^{\alpha}-u^{\alpha}}{u^{\alpha}(s-u)},
  \label{eq:10}
\end{equation}
while Eq.~\eqref{eq:chi_lapl} becomes \cite{aging}
\begin{equation}
  \chi_{t_a,t}(r\vert c) = \delta(r)[1-m_c(t_a,t)]+h_c(t_a,t)\ast_{t}
  \chi_{t}(r\vert c) 
  \label{eq:11}
\end{equation}
where $m_c(t_a,t)=\int_{0}^{t}h_c(t_a,t')dt'$, the symbol $\ast_t$ means
convolution with respect to the variable $t$, and
$\chi_t(r \vert c) = \chi_{t_a=0, t}(r \vert c)$. For $t_a>0$, we
observe an increasing probability of counting exactly $r=0$ events
during the time interval $[t_a,t_a+t]$, with a relative weight given by
$1-m_c$, which will have an important impact in the shape of the degree
distribution.

\subsection{Slightly aged regime}

We expect different aging effects according to the relative importance
of the aging time $t_a$ and the observation time window $t$.  For
slightly aged networks, in which $1\ll t_a \ll t$, Eq.~\eqref{eq:10}
reduces to
\begin{equation}
  h_c(u,s) \simeq \dfrac{1}{u} - \dfrac{s^{\alpha}}{u^{\alpha+1}},
\end{equation}
and Eq.~\eqref{eq:11} is expressed as \cite{aging}
\begin{equation}
  \chi_{t_a,t}(r\vert c) \simeq
  \dfrac{\delta(r)(t_a/t)^{\alpha}}{\Gamma_{1+\alpha}\Gamma_{1-\alpha}}+
  \chi_{t}  +\dfrac{(c\,t_a)^{\alpha}}{\Gamma_{1+\alpha}}\dfrac{\partial
    \chi_{t}}{\partial r} ,
\end{equation}
where we write $\chi_t \equiv \chi_{t}(r\vert c)$ for brevity.
Inserting this expression into Eq.~\eqref{eq:pk} we obtain
\begin{eqnarray}
\label{eq:43}
  P_{t_a,t}(k)&=&\sum_{c}\eta(c)\int_{0}^{\infty}
          \left(\dfrac{(c\,t_a)^{\alpha}}{\Gamma_{1+\alpha}}\dfrac{\partial  
          \chi_{t}}{\partial r}+\chi_{t}\right)g(k\vert r)dr
          \nonumber\\ 
&+&\mathcal{P}(k,\av{r})\dfrac{(t_a/t)^{\alpha}}{\Gamma_{1+\alpha}\Gamma_{1-\alpha}}
\end{eqnarray}
where $\mathcal{P}(k,\av{r})$ is a Poisson distribution centered at
$\av{r}$.  Noticing that for $\av{r} \gg 1$, the Poissonian propagator
in Eq.~\eqref{eq:31} tends to a Gaussian distribution and is as such a
quasi-symmetric function with respect to the axis $k=r+\av{r}$, we write
$g(k\vert r) \simeq g(k-r-\av{r})$.  
Moreover, the dependence of $g$ on the aging time is fully included in 
the average number of activation $\av{r} \equiv \av{r}_{t_a, t}$, 
thus we can use the following relation between $g_0=g_{0,t}$ and $g=g_{t_a,t}$:
\begin{equation}
 g(k\vert r)\simeq g(k_a-r-\av{r}_{0,t})\simeq g_0(k_a\vert r),
\end{equation}
where $k_a=k+\av{r}_{0,t}-\av{r}_{t_a,t}$. 
Inserting this result in Eq.~\eqref{eq:43}, and integrating by parts, we obtain
\begin{equation}
  P(k)\simeq P_0(k_a)+\dfrac{t_a^{\alpha}}{\Gamma_{1+\alpha}}\sum_c
  \eta(c) c^{\alpha}[ P_0(k_a\vert c)-P_0(k_a-1\vert c)] , \nonumber
\end{equation}
where $P_0(k\vert c)=\sum_r \chi_t(r\vert c)g_0(k\vert r)$ is the
non-aged degree distribution for a constant activity $c$.  
This gives, for a distribution $\eta(c)$ with a power-law form given by Eq.~\eqref{eq:eta_c},
\begin{equation}
  P(k)\simeq P_0(k_a)+\dfrac{\beta \,
    t_a^{\alpha}}{(\beta-\alpha)\Gamma_{1+\alpha}}[
  \widetilde{P}_0(\tilde{k}_a)-\widetilde{P}_0(\tilde{k}_a-1)] ,
\label{eq:28}
\end{equation}
where we have defined
$\widetilde{P}_0(k)=\sum_c \tilde{\eta}(c)P_0(k\vert c)$ as the non-aged
degree distribution with a modified activity distribution $\tilde{\eta}$
of parameter $\tilde{\beta}=\beta-\alpha$, and
$\tilde{k}_a=k+\av{r}_{0,t}^{(\tilde{\eta})}-\av{r}_{t_a,t}$, where
$\av{r}_{0,t}^{(\tilde{\eta})}=\sum_c \tilde{\eta}(c)\sum_r
r\,\chi_t(r\vert c)$.  Thus, at large degree and leading order in
$t_a/t$, the second term of the equation is negligible and the aged
degree distribution is simply equal to the non-aged distribution $P_0$
evaluated at $k=k_a$
\begin{equation}
  P_{t_a,t}(k) \sim (c_0t)^{\beta}(k-\av{r}_{t_a,t})^{-\gamma}.
  \label{eq:29}
\end{equation}
Unsurprisingly, the degree distribution $P(k)$ of the slightly aged
network exhibits the same scaling behavior as that of the non-aged one
at large $k$, and we recover the expected expression for a vanishing
$t_a$.  Interestingly, in Eq.\eqref{eq:28} the aged degree distribution
is expressed with two non-aged distributions $P_0$ and
$\widetilde{P}_0$, and the dependence on the aging time $t_a$ is
entirely embedded in the shifted degree $k_a$ and a scaling factor
$t_{a}^{\alpha}$. This allows for a direct evaluation of the aged
distribution, whatever $t_a$, with the prior knowledge of $P_0$ and
$\widetilde{P}_0$ only. In practice however, those two functions are
evaluated via numerical simulations.

\begin{figure}[t]
  \includegraphics[width=\columnwidth]{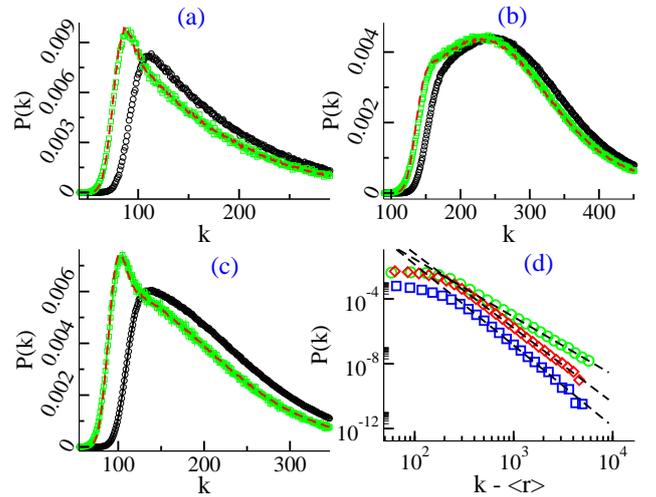}
  \caption{Slightly aged degree distribution $P(k)$ for different values
    of $\alpha$, $\beta$ $t_a$ and $t$.  Plots $(a)$, $(b)$ and $(c)$ show the
    non-aged distribution in black circles and the aged distribution in
    green squares.  The behavior predicted by Eq.~\eqref{eq:28} (with
    $P_0$ and $\widetilde{P}_0$ previously calculated numerically) is
    plotted in red dashed line.  Plot $(d)$ shows the power law behavior
    at large $k$ for the three aged distributions shown in the other
    plots: $(a)$ squares, $(b)$ circles, and $(c)$ diamonds.
    Eq.~\eqref{eq:29} is plotted as a dashed line.  Network size
    $N=10^7$, results are averaged over 50 runs.  The values of the
    parameters are the following.  $(a)$: $(\alpha,\beta)=(0.3,1.2)$,
    $t=10^6$ and $t_a=10^3$.  $(b)$: $(\alpha,\beta)=(0.7,1.8)$, $t=500$
    and $t_a=10$.  $(c)$: $(\alpha,\beta)=(0.5,1.8)$, $t=5.10^3$ and
    $t_a=100$. }
  \label{fig:agedDistr}
\end{figure}

Fig.~\ref{fig:agedDistr} checks the previous results by means of numerical simulations of the NoPAD model 
in the slightly aged regime. We numerically estimate the distributions $P_0$, $\widetilde{P}_0$ and $P$ 
for a network of size $N=10^7$, with three different sets of parameters $(\alpha,\beta,t_a,t)$.  
For each case we compare the aged degree distribution, the non-aged degree distribution and 
the degree distribution given by Eq.~\eqref{eq:28}. 
 One can see that that Eq.~\eqref{eq:28} nicely predicts the aged degree distribution.
Moreover, we observe a bump in the aged degree distribution for small degree values 
with respect to the non-aged distribution, more or less visible depending on the aging time $t_a$.  
The fact that more individuals have a smaller degree in the aged networks means that the
dynamics in this case is slowed down with respect to the non-aged case.
Panel $(d)$ of Fig.~\ref{fig:agedDistr} confirms that the exponent of
the power law decay, $\gamma = 1+\beta/\alpha$, predicted by Eq.~\eqref{eq:29}, is correct.

\subsection{Strongly aged regime}

\begin{figure}[t]
  \includegraphics[width=\columnwidth]{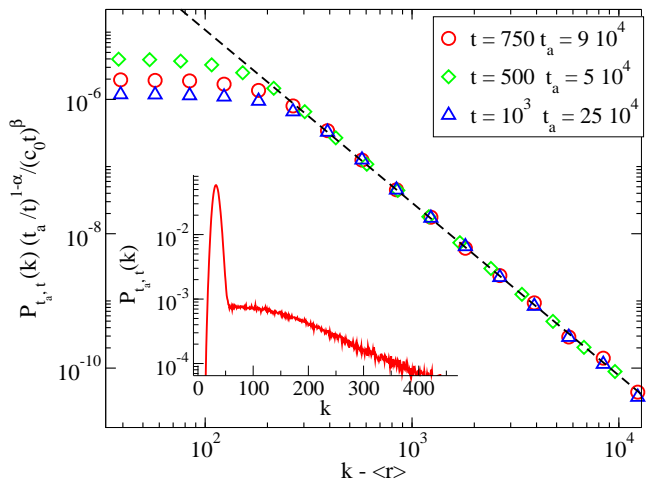}
  \caption{Rescaled degree distribution $P_{t_a,t}(k)$ in case of strong
    aging.  Different values of the time window $t$ and the aging time
    $t_a$ are shown.  Inset: Poissonian behavior of the degree
    distribution $P(k)$ for small $k$ with $t_a=10^5$, $t=500$.  Network
    size $N=10^7$. Parameters are set at $\alpha=0.7$ and $\beta=1.1$.}
  \label{fig:highlyAgedDistr}
\end{figure}

The strongly aged network regime emerges for $1 \ll t \ll t_a$.  In this
limit, the forward waiting in the Laplace space can be approximated as
\begin{equation}
  h_c(u,s) \simeq \dfrac{s^{\alpha-1}}{u^{\alpha}},
\end{equation}
and the aged activation distribution is given by
\begin{equation}
  \chi_{t_a,t}(r\vert c) \simeq
  \left(1-\dfrac{(t/t_a)^{1-\alpha}}{\Gamma_{\alpha}\Gamma_{2-\alpha}}\right)\delta(r)
  - \dfrac{c^{\alpha}
    t_a^{\alpha-1}}{\Gamma_{\alpha}}\int_0^{t}\dfrac{\partial
    \chi_{t'}}{\partial r}dt'. 
    \nonumber
\end{equation}
Using the same approximations as in the slightly aged case, we find,
for $\eta(c)$ given by Eq.~\eqref{eq:eta_c},
\begin{equation}
  \label{eq:pk_aged_poisson}
  P(k) \simeq
  \mathcal{P}(k,\av{r})-\dfrac{\beta}{\tilde{\beta}\,
    \Gamma_{\alpha}t_a^{1-\alpha}} \int_0^{t}dt'[
  \widetilde{P}_{0,t'}(\tilde{k'}_a)-\widetilde{P}_{0,t'}(\tilde{k'}_a-1)] ,
\end{equation}
where $\tilde{k'}_a=k+\av{r}_{0,t'}^{(\tilde{\eta})}-\av{r}_{t_a,t}$.
This expression shows the presence of a population splitting: A majority
of individuals remain inactive over the whole observation time window
$t$, while they still receive connections from the active part of the
population.  This leads to a dominant Poisson term in the degree
distribution.  Again, we find a power law behavior at large $k$,
\begin{equation}
  P_{t_a, t}(k)\sim  (c_0t)^{\beta}(t/t_a)^{1-\alpha}(k-\av{r})^{-\gamma}
\label{eq:pk_aged}
\end{equation}
but this time the tail of the distribution vanishes when $t_a$ tends to
infinity.  Fig.~\ref{fig:highlyAgedDistr} shows the validity of the
scaling with the aging time $t_a$ predicted by Eq.~\eqref{eq:pk_aged}
and the Poissonian term highlighted in Eq.~\eqref{eq:pk_aged_poisson}.

\section{Percolation dynamics}
\label{sec:perco}

Among the topological properties of the integrated networks as a
function of the time, a particularly relevant one is the birth and
evolution of a giant connected component, which constitutes a
percolation process \cite{Newman2010}.  As time passes, more
connections will be established in the integrated network, forming a
growing connected component until at some time $T_p$ this component will
percolate, i.e. it will have a size proportional to the network size
$N$.  The percolation threshold $T_p$ is particularly relevant for the
evolution of dynamical processes running on top of the underlying
network \cite{barratbook}, since any process with a characteristic
lifetime $\tau < T_p$ will be unable to explore a sizable fraction of
the network.

\subsection{General case}

In order to find an expression for the percolation threshold, we will
follow the general formalism valid for correlated random networks, where
the effect of degree correlatons are accounted for by the branching
matrix \cite{PhysRevE.78.051105,citeulike:12856686}
\begin{eqnarray}
  B_{kk'}(t_a,t) =(k'-1)P_{t_a,t}(k'\vert k),
\end{eqnarray}
which implicitly depends on the aging time $t_a$ and the observation
window $t$ through the conditional probability $P_{t_a,t}(k'\vert k)$
that a node with degree $k$ is connected to a node with degree $k'$, in
the time window $[t_a, t_a+t]$ \cite{alexei}.  The percolation threshold
is determined by the largest eigenvalue $\Lambda(t_a,t)$ of the
branching matrix $B_{kk'}(t_a,t)$, which following
\cite{citeulike:12856686} can be written as
\begin{equation}
  \Lambda(t_a,t) = \dfrac{\av{k}}{2}+\dfrac{1}{2}
  \sqrt{4\av{k^2}-4\av{k}-3\av{k}^2},
\end{equation}
where the first and second moment of the degree distribution are
computed on the network integrated in the time window $[t_a, t_a+t]$.
One can express $\Lambda$ as a function of $\av{r} $ and $\av{r^2}$ by
combining Eqs.~\eqref{eq:pk} and~\eqref{eq:gk}, using the hidden
variables formalism \cite{PhysRevE.68.036112}, as
\begin{equation}
  \Lambda(t_a,t)=\av{r}_{t_a,t}+\sqrt{\av{r^2}_{t_a,t}-\av{r}_{t_a,t}}.
\end{equation}
The percolation time $T_p$ determined by imposing the condition
$\Lambda(t_a, T_p) = 1$ \cite{citeulike:12856686}, is thus given by the
solution $T_p$ of the implicit equation
\begin{equation}
\av{r^2}_{t_a,T_p} - \av{r}_{t_a,T_p}^2  = 1-\av{r}_{t_a,T_p}.
\label{eq:21}
\end{equation}
It is worth noting that this result is valid regardless of the age of
the network. However, no explicit expressions exist for $\av{r}_{t_a,t}$
and $\av{r^2}_{t_a,t}$, except for an exponential waiting time
distribution (Poisson process).  Since the network percolation occurs at
relatively short times (such that $\av{r} < 1$), the approximations for
$\chi_t$ performed in the previous Sections cannot be applied, and one
must resort in principle to numerical simulations to estimate $T_p$.

\subsection{Non-aged networks}

In the following, we will study the percolation time $T_p$ for a NoPAD
network with an inter-event time distribution of the form given by
Eq.~\eqref{eq:14} and a parameter $c$ distributed according to
Eq.~\eqref{eq:eta_c} in non-aged networks with $t_a = 0$.

\begin{figure}[t]
  \includegraphics[width=\columnwidth]{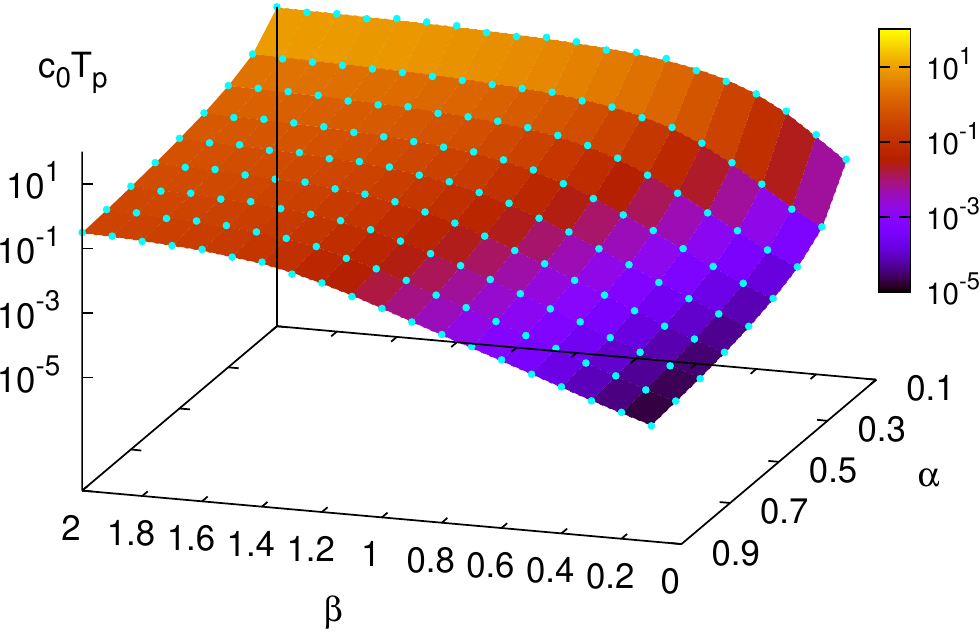}
  \caption{Percolation threshold $T_p$ as a function of $\alpha$ and
    $\beta$.  Blue dots correspond to estimations of $T_p$ as given by
    the peak of the clusters' susceptibility; the surface is obtained by
    a numerical solution of Eq.\eqref{eq:21}.  Network size is
    $N = 10^8$, $c_0=1$ and $c_{max}=10^6$ (see main text). }
  \label{fig:perc6}
\end{figure}

Fig.~\ref{fig:perc6} shows the percolation time $T_p$ as a function of
the two main parameters of the NoPAD model, $\alpha$ and $\beta$,
computed by solving numerically Eq.~\eqref{eq:21}, by means of a
dichotomic search, with $\av{r^2}_{t_a,T_p}$ and $\av{r}_{t_a,T_p}$
evaluated from numerical simulations. We proceed as follows: We evaluate
$\theta_0=1-\av{r}_{t_0}- \av{r^2}_{t_0} + \av{r}_{t_0}^2$ at some
arbitrary starting time $t_0$, and then calculate $\theta_1$ at
$t_1=t_0 \times 2^{\mathrm{sgn}(\theta_0)}$, where $\mathrm{sgn}(z)$ is
the sign function, and so on recursively,
$\theta_k=1-\av{r}_{t_k}- \av{r^2}_{t_k} + \av{r}_{t_k}^2$, with
$t_{k+1}=t_{k} \times 2^{\mathrm{sgn}(\theta_{k})}$.  As $\theta_k$ has
positive values below $T_p$ and negative values beyond, $t_k$ rapidly
converges towards $T_p$. We then repeat the process several times for
decreasing values of the common ratio of the progression until a
predefined precision is reached.  Fig.~\ref{fig:perc6} contrasts the
result of this numerical evaluation of Eq.~\eqref{eq:21} with
estimations of the threshold $T_p$ from numerical simulations of the
NoPAD model, defined by means of the peak of the clusters'
susceptibility $\chi(s)$ \cite{stauffer94,citeulike:12856686}.  The
susceptibility $\chi(s)$ is defined as $\chi(s)=\sum_s s^2\,n_s$, where
$n_s$ is the density of clusters of size $s$ and the sum is restricted
to all clusters, except the largest one.  Fig. ~\ref{fig:perc6} shows
that the two methods are in very good agreement, although the threshold
obtained by means of $\chi(s)$ tends to be slightly above to the one
predicted by Eq.~\eqref{eq:21}, with an average error of $6\%$.

\begin{figure}[t]
  \includegraphics[width=\columnwidth]{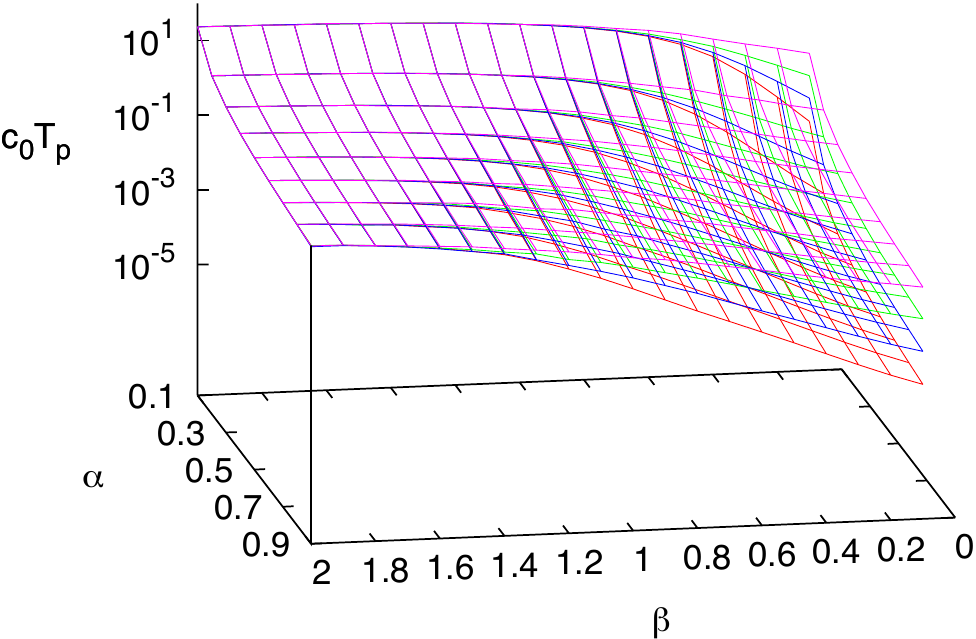}
  \caption{Percolation threshold $T_p$ as a function of $\alpha$ and
    $\beta$ for different values of $c_{max}$.  $T_p$ is calculated
    numerically from Eq.~\eqref{eq:21}, with $c_0=1$.  From top surface
    to bottom one, values of $c_{max}= 10^3,10^4,10^5,10^6$.  Network
    size is $N = 10^8$. }
  \label{fig:cmaxvarying}
\end{figure}

One can see that the percolation time $T_p$ rapidly decreases toward
zero in a region of the ($\beta, \alpha$) space.  This is due to the
fact that, as Eq.~\eqref{eq:21} shows, if the second moment $\av{r^2}_t$
diverges, then $T_p$ tends to zero in the thermodynamic limit,
$N \gg 1$.  For a non-aged agent with activity $c$, one has, for
$ct \gg 1$ \cite{renewal}
\begin{equation}
\overline{r^2}_t(c) = \sum_r r\,\chi_t(r\vert c) \sim (ct)^{2\alpha}.
\end{equation}
Thus for $\beta < 2\alpha$,
$\av{r^2}_t = \int dc \; \eta(c) \overline{r^2}_t(c)$ is a divergent
function at large times.  This implies that $\av{r^2}$ is infinite
$\forall t > 0$, since otherwise there would be a discontinuity at some
arbitrary time $t>0$, which makes no sense.  Therefore, the percolation
time is zero in the thermodynamic limit for $\beta < 2\alpha$, while it
is finite otherwise.  To avoid these finite size effects, one need to
set a cutoff $c_{max}$ for the parameter $c$ (in Fig.~\ref{fig:perc6}
this cutoff is set to $c_{max} = 10^6$).  We explore the impact of the
cutoff in Fig.~\ref{fig:cmaxvarying}, which shows the percolation time
$T_p$ in the ($\beta, \alpha$) space for different values of $c_{max}$.
We choose a network size $N$ such that $N\geq 100\,c_{max}$ in order to
hinder sampling errors on the values of $c$.  As expected, we observe a
strong decay of $T_p$ towards zero in the region $\beta<2\alpha$, as
$c_{max}$ grows.

\begin{figure}[t]
  \includegraphics[width=\columnwidth]{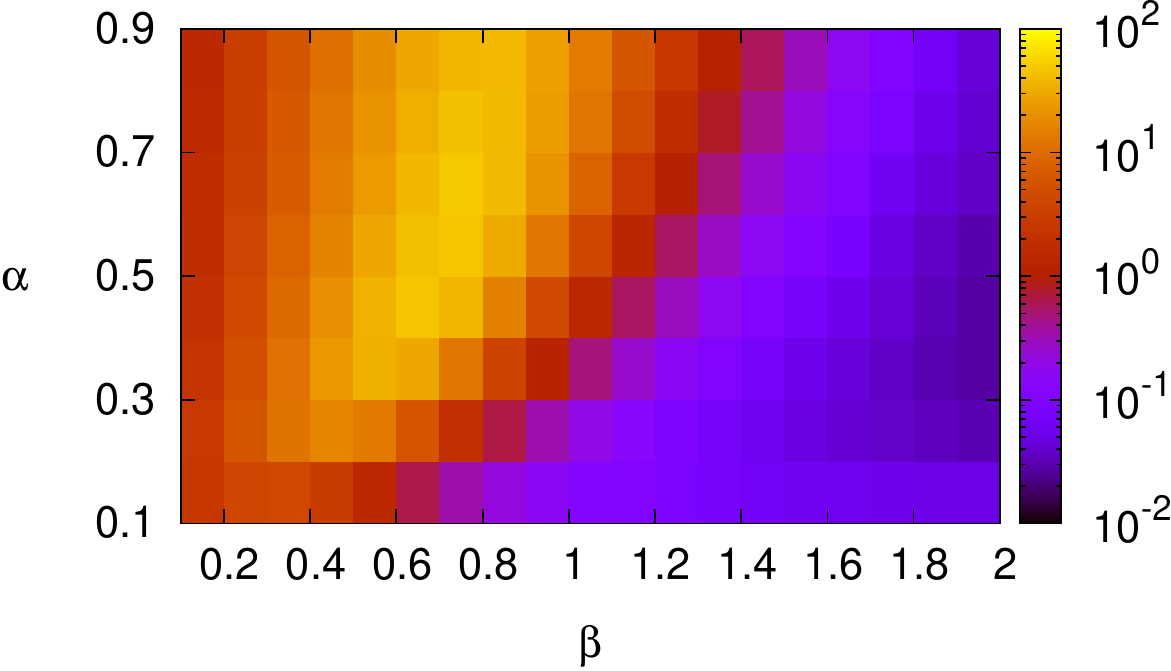}
  \caption{Relative error $(T_{p}-T_{p}^{0})/T_{p}^0$ as a function of
    $\alpha$ and $\beta$.  $T_p^0$ is obtained by numerically solving
    the implicit equation ensuing from tthe Molloy-Reed criterion, $T_p$
    is given by the numerical solution of Eq.~\eqref{eq:21}.  Network
    size is $N=10^8$, $c_0=1$ and $c_{max}=10^6$. }
  \label{fig:criterionError}
\end{figure}

We also compare the percolation threshold obtained within the correlated
networks formalism, $T_p$, with the prediction valid for uncorrelated
networks, as given by the Molloy-Reed (MR) criterion \cite{molloy98}.  The
MR criterion imposes the presence of a giant component whenever the
condition $\av{k^2} / \av{k} > 2$ is fulfilled, which in the present
case translates in a percolation time $T_p^0$ given by the solution of
the equation $3\av{r}^2_{T_p^0}+\av{r^2}_{T_p^0}-3\av{r}_{T_p^0}=0$,
which can be solved numerically applying the dichotomic search method
described above.  Fig.~\ref{fig:criterionError} shows the relative error
between $T_p$ and $T_p^0$, in the ($\beta, \alpha$) space.  One can see
that only for large values of $\beta$ the MR criterion is close to the
real percolation threshold, justifying the necessity of using the
correlated networks formalism.

\begin{figure}[t]
  \includegraphics[width=0.9\columnwidth]{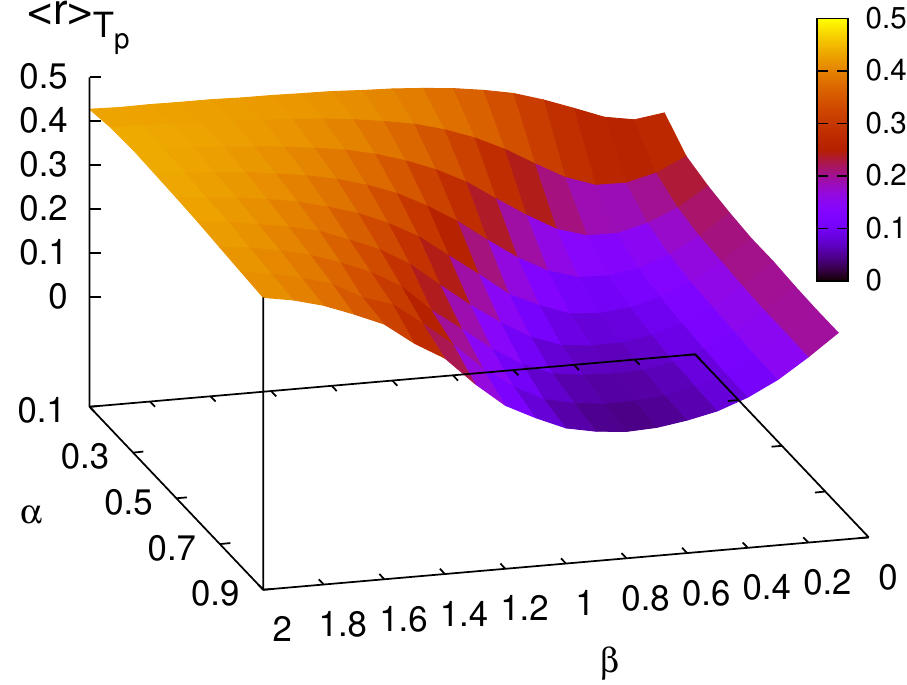}
  \caption{Average activation number at the threshold, $\av{r}_{T_p}$,
    as a function of $\alpha$ and $\beta$ Network size $N=10^7$, $c_0=1$
    and $c_{max}=10^6$.}
  \label{fig:r_avg}
\end{figure}

Another interesting observation comes from relating the behavior of the
percolation time $T_p$ as a function of $\alpha$ and $\beta$, shown in
Fig.~\ref{fig:perc6}, with the average number of activation events
counted in the time window $[0,T_p]$, $\av{r}_{T_p}$, which is a measure
of the density (average degree) of the integrated network.  On the one
hand, increasing $\beta$ while keeping constant $\alpha$ decreases
$\av{r}_{T_p}$, and so it increases the percolation threshold $T_p$.  On
the other hand, increasing $\alpha$ while keeping $\beta$ constant
accelerates the growth of the integrated network, so $\av{r}_{T_p}$
increases and $T_p$ is smaller.  The average number of activations
$\av{r}_{T_p}$, thus, as a measure of the density of the network at time
$t=T_p$, provides useful additional information on the characteristics
of the percolation process.  Fig.~\ref{fig:r_avg} displays
$\av{r}_{T_p}$ as a function of $\alpha$ and $\beta$, showing that this
density has a minimum value which appears to be close to the region
$\alpha=\beta$.  In this region, agents form a giant component even
though they hardly have interacted, indicating that the link emission
pattern is more efficient for this particular set of parameters.  A
partial explanation of this feature can be derived from an evaluation of
Eq.~\eqref{eq:21} in the large time limit, even though the network
percolates at times where asymptotic expansions of $\av{r}_t$ and
$\av{r^2}_t$ are not relevant.  In this limit we write
$\av{r^n}\simeq \dfrac{\Gamma_{n+1}}{\Gamma_{\alpha n
    +1}}\,\av{c^{\alpha n}}t^{\alpha n}$ \cite{aging}, which gives an
average activation number at the threshold $T_p$
\begin{equation}
  \av{r}_{T_p} \simeq
  \dfrac{\sqrt{1+4R(\alpha,\beta)}-1}{2R(\alpha,\beta)},
  \label{eq:12}
\end{equation} 
where
$R(\alpha,\beta)=2\,\Gamma_{\alpha+1}^2 \av{c^{2\alpha}} /
\Gamma_{2\alpha+1} \av{c^{\alpha}}^2 - 1$. The possible extremes in Eq.~(\ref{eq:12}), for a given value of $\alpha$, correspond to
solution of $\partial R(\alpha,\beta) / \partial \beta = 0$. Performing
the integrals in the definition of $R(\alpha,\beta)$ (with a maximum
$c_\mathrm{max}$ to avoid divergences) and taking the partial derivative
with respect to $\beta$ leads to a minimum in $\av{r}_{T_p}$ located
precisely at $\beta = \alpha$, in qualitative agreement with Fig.~\ref{fig:r_avg}.

As stated above, for a the general form of the waiting time distribution
given by Eq.~\eqref{eq:14}, neither explicit expressions are available
for $\av{r}_{T_p}$ and $\av{r^2}_{T_p}$, nor are approximations valid
close to the percolation time $T_p$, so that one must resort to
numerical simulations to estimate $T_p$.  An exception is the case of a
power law waiting time distribution with exponent $\alpha=1/2$, which
corresponds to the one-sided L\'evy distribution
\cite{klafter_first_2011}
\begin{equation}
  \psi_c(t) = \frac{e^{-1/(ct)}}{\sqrt{\pi c}\,\, t^{3/2}}.
\end{equation}
In this case, the distribution of activation numbers at time $t$ reduces
to
\begin{equation}
  \chi_t(r\vert c) =
  \mathrm{erf}\left(\dfrac{r+1}{\sqrt{ct}}\right)-\mathrm{erf}
  \left(\dfrac{r}{\sqrt{ct}}\right)  ,
\end{equation}
where $\textrm{erf}(z)$ is the error function. The moments of the
activation distribution can be analytically expressed as
\begin{equation}
  \av{r^n}_t = \int dc \,\eta(c) \sum_{r=0}^{\infty} r^n
  \left(\mathrm{erf}\left(\dfrac{r+1}{\sqrt{ct}}\right)-\mathrm{erf}
    \left(\dfrac{r}{\sqrt{ct}}\right)\right)  ,
\label{eq:r_n}
\end{equation}
and the percolation time $T_p$ can be computed by introducing
Eq.~(\ref{eq:r_n}) into Eq.~(\ref{eq:21}) and solving numerically the
ensuing self-consistent equation.

\begin{figure}[t]
  \includegraphics[width=\columnwidth]{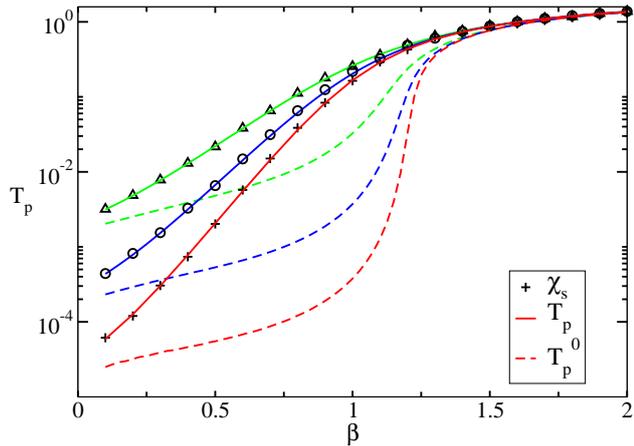}
  \caption{Percolation threshold $T_p$ on a L\'evy NoPAD network as a
    function of $\beta$.  Three different values of the parameter
    $c_{max}$ are shown.  Symbols represent $T_p$ evaluated by means of
    the peak of the susceptibility $\chi(s)$.  Continuous lines
    represent $T_p$ evaluated numerically from Eq.~(\ref{eq:r_n}) and
    Eq.~(\ref{eq:21}).  Dashed lines represent the threshold $T_p^0$
    given by the Molloy-Reed criterion.  Triangles correspond to
    $c_{max}=10^4$, circles to $c_{max}=10^5$ and crosses to
    $c_{max}=10^6$.  The corresponding dashed lines follow the same
    downward progression.  Lower bound activity $c_0=1$.  Network size
    $N=10^8$.}
  \label{fig:levyperc}
\end{figure}

Fig.~\ref{fig:levyperc} shows the percolation time $T_p$ on a L\'evy
NoPAD network as a function of the activity distribution exponent
$\beta$.  One can see that the theoretical prediction fits very well the
numerical estimation of $T_p$ given by the peak of the cluster
susceptibility $\chi(s)$.  In the same Fig.~\ref{fig:levyperc}, we also
plot the percolation threshold $T_p^0$ as predicted by the MR criterion,
showing that this is a good approximation only if $\beta$ is close to
$2$, in accordance with what is observed in
Fig.~\ref{fig:criterionError}, for an inter-event time distribution with
a general form given by Eq.~\eqref{eq:14}.

\subsection{Aged networks}

Here we consider the effects of aging on the percolation threshold.
Eq.~(\ref{eq:21}) is valid also in presence of aging, so that one can
numerically solve it by means of the dichotomic search explained above,
and find the percolation threshold $T_p$.  We checked that this method
works also in presence of aging, $t_a>0$.  However, one can determine
the asymptotic behavior of the percolation threshold $T_p$ as a function
of the aging time $t_a$, in the limit of large aging, as follows.  Since
the main effect of aging is to delay the growth of the integrated
network, we expect the same consequences for the birth of the giant
component.  Therefore, $T_p$ must be an increasing function of $t_a$.
Three different asymptotic behaviors are to be considered when $t_a$
tends to infinity: $T_p/t_a$ either tends to $0$, to a positive constant, or it diverges.
It is straightforward to discard the latter, since in this case one
would have $\av{r}_{t_a,T_p}\simeq \av{r}_{0,T_p} \gg 1$, which is
contradictory with the condition $\av{r}_{T_p}<1$.  Thus, one can look
for a solution satisfying $T_p/t_a \to 0$ for $t_a \to \infty$, and if a
solution is found, then it is the correct one, since it is a lower bound
for any other. Using the expansions for the strong aging regime proposed
in \cite{aging}, for $t_a \gg T_p \gg 1$ one has
\begin{equation}
  \av{r^n}_{t_a, T_p}\simeq \dfrac{\Gamma_{n+1}}{\Gamma_\alpha
    \Gamma_{\alpha n + 2 - \alpha}}\av{c^{\alpha n}}\,t_a^{\alpha-1}\,
  T_p^{1-\alpha+\alpha 
    n}. 
    \nonumber
\end{equation}
By inserting the moments of $r$ in Eq. \eqref{eq:21}, one finds
\begin{equation}
  T_p   \simeq A(\alpha, \beta)  \,\, t_a^{\frac{1-\alpha}{1+\alpha}},
\label{eq:30}
\end{equation}
where $A(\alpha, \beta) = [\Gamma_\alpha \Gamma_{2+\alpha} / 2\av{c^{2\alpha}}
] ^{\frac{1}{1+\alpha}}$.

\begin{figure}[t]
  \includegraphics[width=\columnwidth]{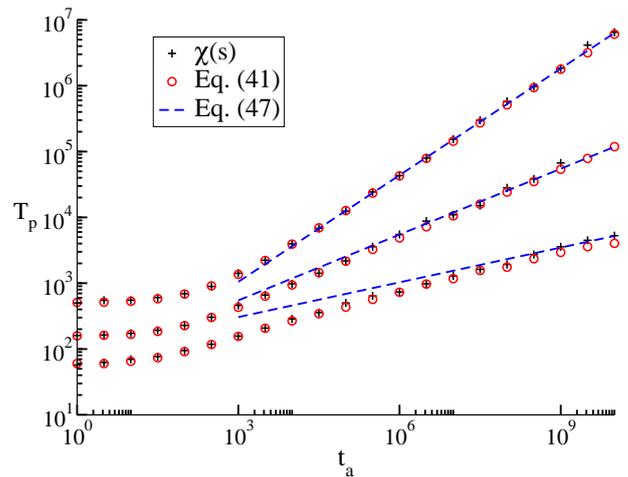}
  \caption{Percolation threshold $T_p$ as a function of the aging time
    $t_a$.  Parameters are set to $\beta = 1.5$, $c_0=0.001$,
    $c_{max}=1$ and $\alpha = 0.3, 0.5, 0.7$ from top to bottom.
    Circles represent $T_p$ evaluated numerically from \eqref{eq:21}.
    Crosses are an estimation of $T_p$ as given by the peak of the
    susceptibility.  The asymptotic behavior predicted by \eqref{eq:30}
    is plotted in dashed line.  Network size $N=10^6$. }
  \label{fig:agedperc}
\end{figure}

Fig.~\ref{fig:agedperc} shows the percolation time $T_p$ evaluated from
Eq. \eqref{eq:21}, using a dichotomic search strategy, as a function of
the aging time $t_a$, and computed from direct numerical simulations
using the susceptibility peak, for a NoPAD network with $\beta=1.5$ and
different values of $\alpha$.  One can observe that aging has
practically no effect on the percolation time $T_p$ for $t_a$ smaller
than the percolation threshold with no aging.  On the contrary, for
$t_a \gg T_p \gg 1$, the asymptotic behavior of $T_p$ as a function of
$t_a$ is very well predicted by Eq. \eqref{eq:30}.

\section{Discussion}
\label{sec:conclusions}

In the study of complex systems, one of the main assets of statistical physics consists in 
the postulation of simple models capable to reproduce one given relevant property of the system under consideration.
This approach allows to simplify the study, by focusing on the property under scrutiny, 
independently of other complicating factors.  
In the case of static complex networks, the configuration model fulfills this role with respect to 
the degree distribution, by considering networks characterized exclusively by this degree distribution, 
and completely random regarding all other properties. 
In the field of temporal networks, the non-Poissonian activity driven (NoPAD) model fills this niche, 
providing a simple model characterized by an arbitrary inter-event time distribution, 
that assume any form, in particular that dictated by empirical evidence.

In this paper we have presented a detailed mathematical study of the properties of 
the time-integrated networks emerging from the dynamics of the NoPAD model.  
We have focused in two main issues: The topological properties of the integrated networks, 
and their percolation behavior, as determined by the percolation time $T_p$ at which 
a giant connected component, spanning a finite fraction of total number of nodes in the network, first emerges. 
 These two properties are determined as a function of the model's parameters, 
 namely the exponent $\alpha$ of the waiting time distribution $\psi_c(t)$, 
 and the exponent $\beta$ of the agents' heterogeneity distribution $\eta(c)$, 
 as well as a function of the time window of the integration process $[t_a, t_a+t]$, 
 by applying a mapping of the network's construction algorithm to the hidden variables class of models.  
 For the case of the degree distribution $P(k)$, we recover the intimate connection between 
 the scale-free nature of static social networks, $P(k) \sim k^{-\gamma}$, 
 and two main characteristics of social temporal networks, 
 namely a power-law distributed waiting time, $\psi_c(t) \sim (t c)^{-1-\alpha}$, 
 and a power-law form of the heterogeneity distribution, $\eta(c) \sim c^{-1-\beta}$, 
 as deduced from the distribution of average activity \cite{2012arXiv1203.5351P,PhysRevLett.114.108701}.  
This relation is quantified in the identity $\gamma = 1+\beta/\alpha$.  
With respect to the percolation time $T_p$, analytic equations are obtained,
from which the value of $T_p$ can be obtained by solving them numerically. 
A relevant result here is that the percolation time vanishes in the thermodynamic limit in the region $\beta < 2\alpha$, 
where the fast aggregation of connections leads to a giant component in a very
short interval of time.

The main asset of the NoPAD model is that it allows to transparently observe the aging effects 
introduced by arbitrary waiting time distributions.  
These effects are due to the agent's memory from his last activation time, 
memory that is always present in renewal processes, with different degrees of severity, 
unless all nodes follow memoryless Poisson processes.  
Aging effects translate in a breaking of the time translation symmetry, 
and induce a dependence of topological observables on the beginning of the integration window $t_a$, 
and are remarkably strong in the region $\alpha<1$, when the average waiting time of any agent is divergent. 
This aging effects, fully described by the mathematical formalism of the NoPAD model, could be easily guessed,
in base of the empirical evidence of a diverging average waiting time,
in terms of the so-called waiting time paradox \cite{burstylambiotte2013}.

The NoPAD model represents a minimal model of temporal networks with long tailed inter-event time distribution.  
As such, it has a wide potential to serve as a synthetic controlled environment to check 
both numerically and analytically several properties of these networks, and
in particular their effect on dynamical processes, in much the same way
as the configuration model has played this role for static networks.
Moreover, due to its simple definition, it can be easily modified to make it more realistic.  
We envision as the more interesting of those improvements the introduction of 
a finite duration for the contacts between nodes and the addition of memory effects 
in the process of selecting neighbors after an activation process \cite{Karsai:2014ab}. 
On the other hand, it is very easy to show \cite{citeulike:12856686} that
the NoPAD integrated networks exhibit dissasortative degree correlations \cite{PhysRevLett.89.208701}, 
at odds with empirical observations in real static social networks. 
Correcting this effect emerges also as an important future objective.

\begin{acknowledgments}
  We acknowledge financial support from the Spanish MINECO, under
  project FIS2013-47282-C2-2, and EC FET-Proactive Project MULTIPLEX
  (Grant No. 317532).  M.S. acknowledges financial support by the James
  S. McDonnell Foundation.  R. P.-S. acknowledges additional financial
  support from ICREA Academia, funded by the Generalitat de Catalunya.
\end{acknowledgments}


%

\end{document}